\documentclass{PoS}

\usepackage[utf8]{inputenc}
\usepackage{url}

\RequirePackage{subfigure}

\title{Discovery potential of light sterile neutrinos with displaced vertices}

\ShortTitle{Sterile neutrinos with displaced vertices}

\author{\speaker{Giovanna Cottin}\thanks{Based on work done in collaboration with Juan Carlos Helo and Martin Hirsch.}\\
        Department of Physics, National Taiwan University, Taipei 10617, Taiwan\\
        E-mail: \email{gcottin@phys.ntu.edu.tw}}

\abstract{Many models of new physics beyond the Standard Model are able to describe massive, long-lived particles with macroscopic decays, which can be reconstructed as displaced vertices inside the inner trackers of the LHC experiments. In addition, the lack of evidence of any new physics at the LHC motivates to perform more unconventional searches, such as looking for displaced vertices. I comment on the 13 TeV LHC reach with a proposed multitrack displaced vertex search strategy to probe light sterile neutrinos. Limits on active-sterile neutrino mixing are presented.}

\FullConference{The 39th International Conference on High Energy Physics (ICHEP2018)\\
		4-11 July, 2018\\
		Seoul, Korea}

\begin{document}

\section{Introduction}

The understanding of the origin of light neutrino masses in the Standard Model (SM) still remains one fundamental open quest in particle physics. In its simplest form (the type I see-saw~\cite{Minkowski:1977sc}), light neutrino masses can be explained by introducing massive, sterile neutrinos $N$. Within this framework, $N$ mixes with the neutrinos in the SM. For small values of the mixing $V_{lN}$ (with $l=e,\mu,\tau$), $N$ can have a large lifetime. Heavy, long-lived neutrinos are predicted in several models of new physics beyond the SM. Examples include the above mentioned SM extended by one sterile neutrino (SST)~\cite{Mohapatra:1979ia,Schechter:1980gr}, and left-right symmetric extensions of the SM (LR)~\cite{Pati:1974yy,Mohapatra:1974gc}, where $N$ production and decay depends mostly on the unknown mass of the new, heavy right-handed gauge boson, $W_{R}$.  In both cases, $N$ can have a masses covering various mass ranges. If its mass is below the electroweak scale ($\lesssim 100$ GeV), it could decay with a characteristic displaced vertex (DV) signature inside the LHC experiments. 

Most LHC searches for sterile neutrinos exploit the Majorana signature of same-sign dileptons and jets~\cite{Keung:1983uu} and the focus is on masses with $m_{N}\gtrsim 100$ GeV. To date, no dedicated public searches exists at the LHC covering $m_{N} \ll 100$ GeV, which is the low mass regime in which searches for displaced vertices are sensitive. Several theory work was proposed in recent years showing the enlarged parameter space of many models that could be probed when targeting a low sterile neutrino mass region with current/proposed displaced strategies. For example, searches for displaced lepton-jets~\cite{Izaguirre:2015pga}, displaced jets~\cite{Nemevsek:2018bbt} and displaced vertices~\cite{Helo:2013esa} were studied. Here we focus on a displaced vertex search that triggers on the prompt lepton coming from the production of $W$ bosons. The search also has optimized cuts in the vertex selection, described in the next section. I refer the reader to Refs.~\cite{Cottin:2018kmq,Cottin:2018nms} for more details.

\section{Displaced vertex search strategy}

The following selections are imposed inspired by the ATLAS multitrack displaced vertex search~\cite{Aaboud:2017iio,Aad:2015rba}:

\begin{enumerate}

\item{One prompt lepton with $p_{\mbox{T}}>25$ GeV.}

\item{Decay position of the DV within 4 and 300 mm.}

\item{Charged tracks from the DV must have $p_{\mbox{T}}>1$ GeV and transverse impact parameter $|d_{0}|>2$ mm.}

\item{Number of selected tracks $N_{\mbox{trk}}>3$. The invariant mass of the DV must be $m_{\mbox{DV}}\geq5$ GeV.}

\end{enumerate}

\section{LHC reach}

\begin{figure*}[htp]
\centering
\subfigure[$e-$mixing]{\includegraphics[width=0.45\columnwidth]{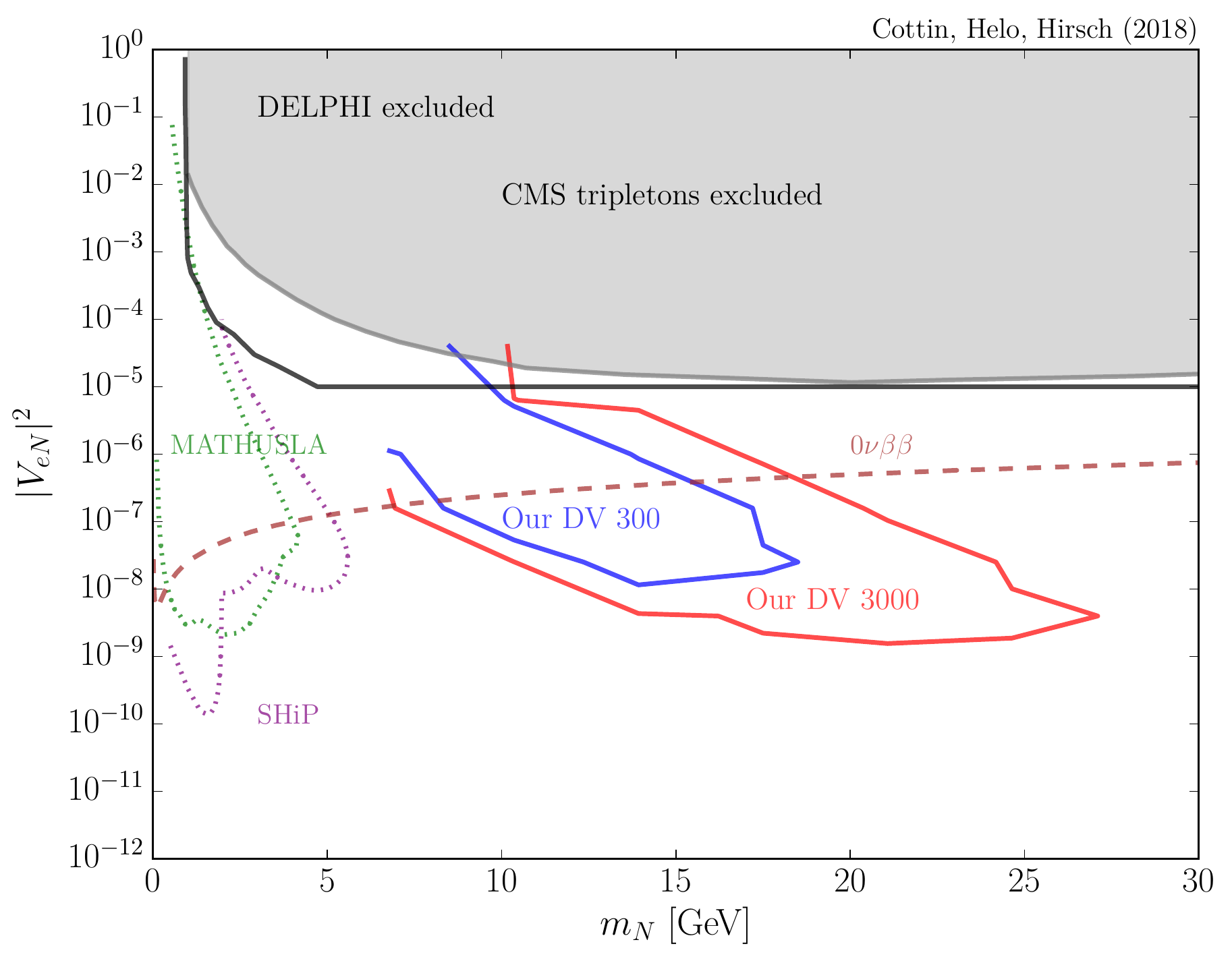}}
\subfigure[$\mu-$mixing]{\includegraphics[width=0.45\columnwidth]{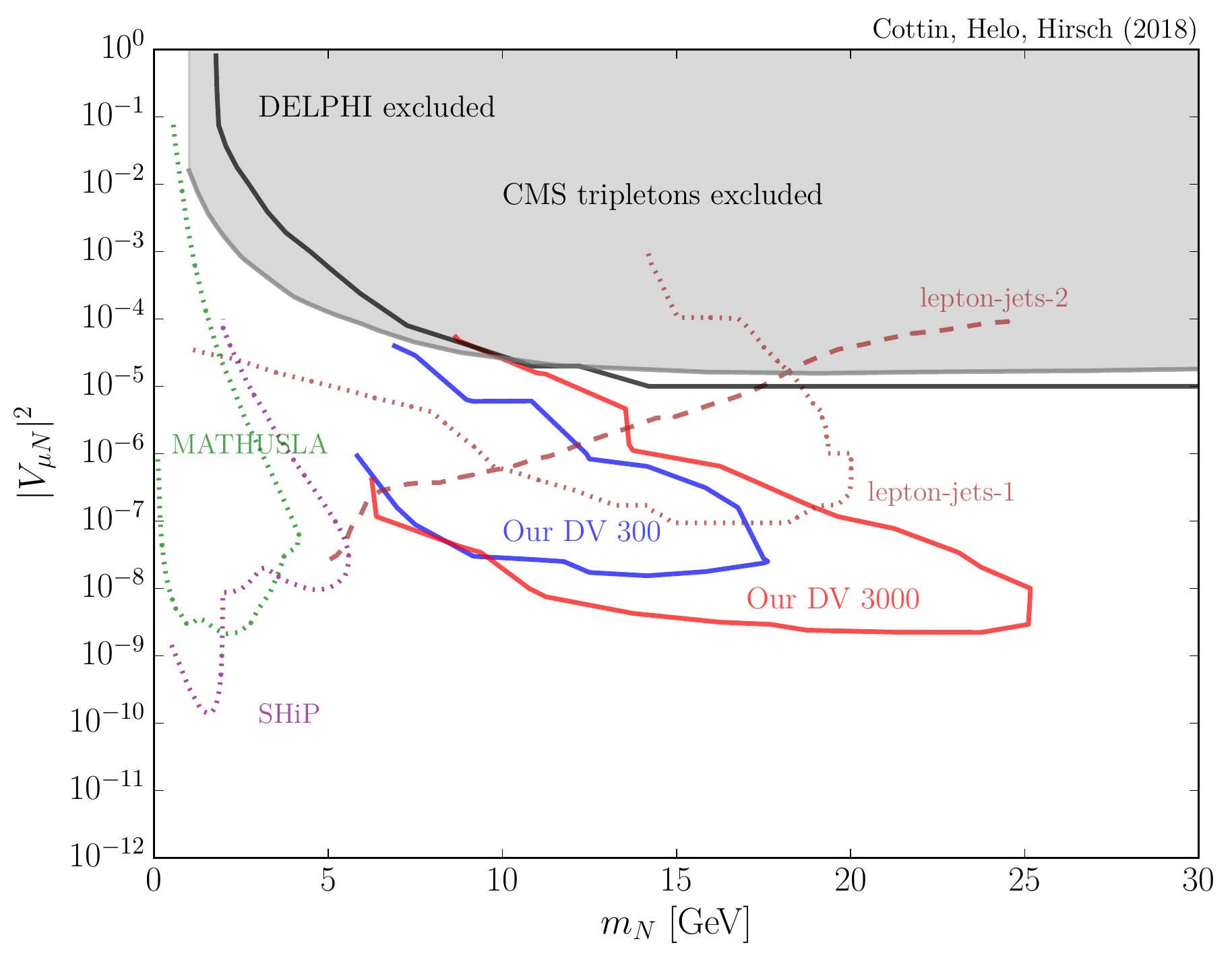}}
\subfigure[$\tau-$mixing]{\includegraphics[width=0.45\columnwidth]{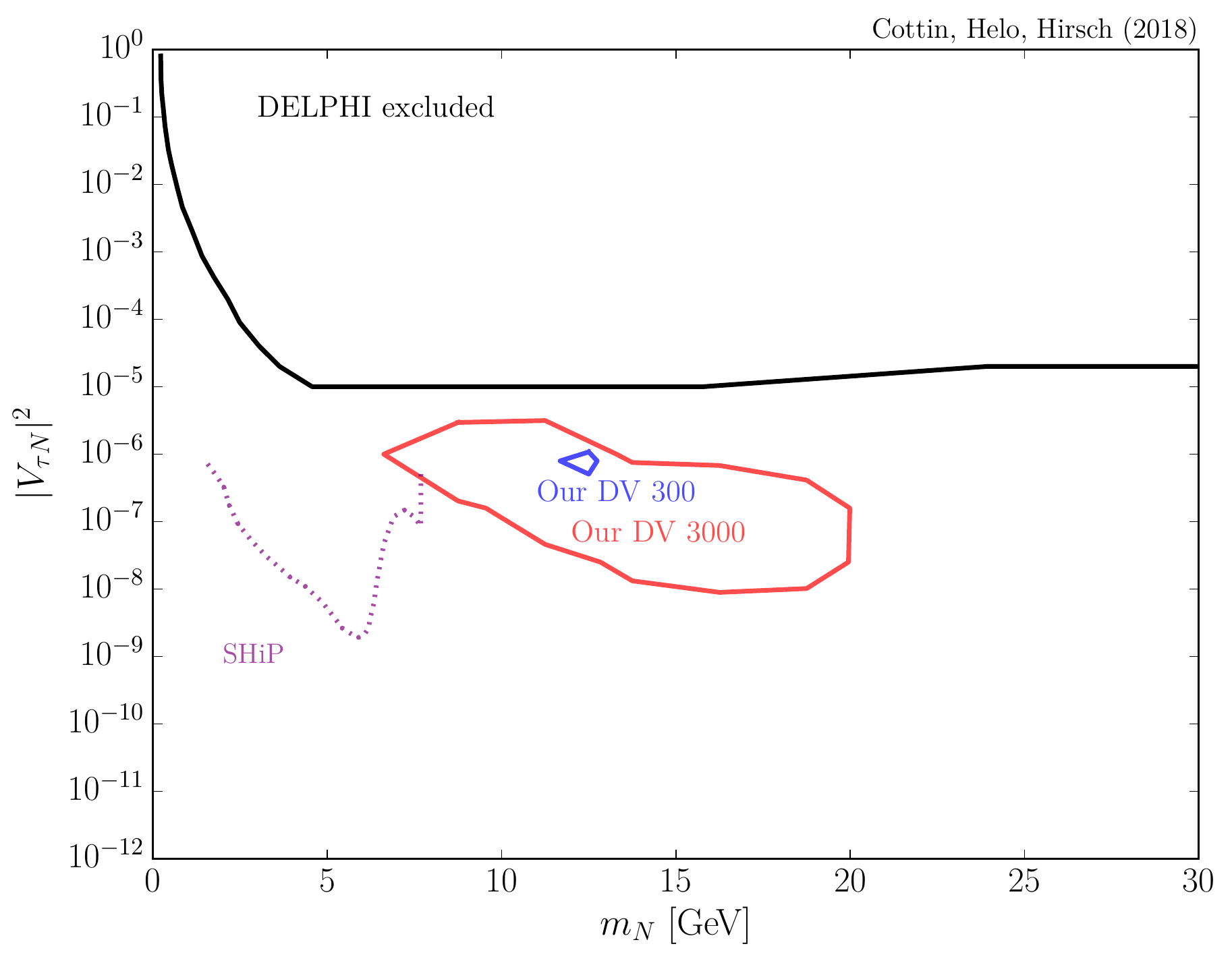}}
\caption{$95\%$ CL reach in the $(|V_{lN}|^2,m_{N})$ plane at
  $\sqrt{s}=13$ TeV of our proposed multitrack displaced strategy for
  $\mathcal{L}=300$ fb$^{-1}$ (blue) and $\mathcal{L}=3000$ fb$^{-1}$
  (red). See Ref.~\cite{Cottin:2018nms} for details on the projected sensitivities shown from other searches/experiments.
\label{fig:LHCreach}}
\end{figure*}

With the above selections, the 13 TeV LHC with $3000$/fb of luminosity can probe $m_{N}\sim 30$ GeV (for a $W_{R}$ boson mass of around 5 TeV) in the LR model~\cite{Cottin:2018kmq}. Figure~\ref{fig:LHCreach} shows the reach in the SST model, separately for mixing in the electron, muon and tau sector. Mixings as low as $|V_{l,N}|^2 \approx 10^{-9}$ can be accessed~\cite{Cottin:2018nms}, showing the best sensitivity achievable to date within $5$ GeV $<m_{N} <30$ GeV with no competition from other experiments in this mass range.

\section{Summary}
Displaced vertex signatures at colliders are well motivated, as they can be a characteristic signal in many frameworks for new physics beyond the Standard Model that benefits from very small backgrounds. 
If sterile neutrinos exist, and if they have masses around or below the electroweak scale, then they will be long-lived particles decaying with spectacular displaced vertices inside the LHC detectors, whose identification can provide a clear collider test of models for neutrino mass generation.

\acknowledgments

I thank the ICHEP 2018 organizing committee for the award of a travel grant and Cheng-Wei Chiang, Kingman Cheung, Pyungwon Ko and NCTS, Taiwan for additional support to attend the conference. I also thank the Korea Institute of Advance Studies (KIAS) for hospitality during my stay in Seoul. This research was supported by the Ministry of Science and Technology (MOST) of Taiwan under Grant No. MOST-106-2811-M-002-035.

\end{document}